\documentclass[showpacs,floatfix,twocolumn,prl]{revtex4-1}

\usepackage{graphicx,color}

\newcommand*{\bto}{Bi$_2$Ti$_2$O$_6$O$^\prime$}
\newcommand*{\bro}{Bi$_2$Ru$_2$O$_6$O$^\prime$}

\newcommand{\obo}{O$^\prime$--Bi--O$^\prime$}

\begin{document}

\title{Incoherent Bi off-centering in Bi$_2$Ti$_2$O$_6$O$^\prime$ and 
Bi$_2$Ru$_2$O$_6$O$^\prime$: Insulator versus metal}

\author{Daniel P. Shoemaker}\email{dshoemaker@anl.gov}
\affiliation{Materials Science Division, Argonne National Laboratory
Argonne, IL, 60439, USA}
\author{Ram Seshadri}\email{seshadri@mrl.ucsb.edu}
\affiliation{Materials Department, University of California, Santa Barbara, CA, 93106, USA}
\author{Makoto Tachibana}\email{tachibana.makoto@nims.go.jp}
\affiliation{National Institute for Materials Science, Namiki 1-1, Tsukuba, 
Ibaraki 305-0044, Japan}
\author{Andrew L. Hector}\email{a.l.hector@soton.ac.uk} 
\affiliation{School of Chemistry, University of Southampton, Highfield, 
Southampton SO17 1BJ, UK} 

%\date{\today}

\begin{abstract}
In the cubic, stoichiometric oxide compounds Bi$_2$Ti$_2$O$_6$O$^\prime$ 
(also written Bi$_2$Ti$_2$O$_7$) and Bi$_2$Ru$_2$O$_6$O$^\prime$ (also written 
Bi$_2$Ru$_2$O$_7$) Bi$^{3+}$ ions on the pyrochlore $A$ site display a 
propensity to off-center. Unlike Bi$_2$Ti$_2$O$_6$O$^\prime$, 
Bi$_2$Ru$_2$O$_6$O$^\prime$ is a metal, so it is of interest to ask whether 
conduction electrons and/or involvement of Bi 6$s$ states at the Fermi energy 
influence Bi$^{3+}$ displacements. The Bi$^{3+}$ off-centering in 
Bi$_2$Ti$_2$O$_6$O$^\prime$ has previously been revealed to be incoherent from
detailed by reverse Monte Carlo analysis of total neutron scattering. 
Similar analysis of Bi$_2$Ru$_2$O$_6$O$^\prime$ reveals incoherent 
off-centering as well, but 
of smaller magnitude and with distinctly different orientational preference. 
Analysis of the distributions of metal to oxygen distances presented suggests
that Bi in both compounds is entirely Bi$^{3+}$. Disorder in 
Bi$_2$Ti$_2$O$_6$O$^\prime$ has the effect of stabilizing valence while 
simultaneously satisfying the steric constraint imposed by the presence of the 
lone pair of electrons. In Bi$_2$Ru$_2$O$_6$O$^\prime$, off-centering is not 
required to satisfy valence, and seems to be driven by the lone pair. Decreased
volume of the lone pair may be a result of partial screening by 
conduction electrons.
\end{abstract}


\maketitle

\section{Introduction} 

The oxide pyrochlores $A_2B_2$O$_6$O$^\prime$, usually abbreviated 
$A_2B_2$O$_7$, are well known for their ability to accommodate magnetic cations 
on interpenetrating sub-lattices of corner-connected O$^\prime A_4$ tetrahedra 
and $B$O$_6$ octahedra. The geometry of the sublattices often result in 
magnetically frustrated ground 
states,\cite{ramirez_strongly_1994,ramirez_zero-point_1999} resulting in 
spin glass, spin ice, or spin liquid phases.\cite{canals_pyrochlore_1998} 
Frustration of concerted atomic displacements (dipolar frustration) 
has also been proposed, when electronic dipoles rather than magnetic spins are 
placed on the pyrochlore $A$ 
site,\cite{melot_displacive_2006,mcqueen,shoemaker_atomic_2010} 
with the suggested ``charge-ice''\cite{seshadri} displaying the appropriate 
entropic signatures.\cite{melot_large_2009} Large atomic displacement 
parameters associated with the $A$-site cation, seen in Fourier maps of 
La$_2$Zr$_2$O$_7$\cite{tabira_annular_2001} and in Bi$_2M_2$O$_7$ 
($B$ = Ti, Zn, Nb, Ru, Sn, Hf),\cite{avdeev_static_2002,hector_synthesis_2004,vanderah_unexpected_2005,henderson_structural_2007} contribute to
the expanding body of evidence that pyrochlores prefer to accommodate cation
off-centering \textit{via} incoherent disorder rather than in ordered non-cubic
ground states. 

Among Bi$_2B_2$O$_7$ pyrochlores, $B$ = Ti, (Zn/Nb), Sn, and Hf are insulators,
while $B$ = Ru, Rh, Ir, and Pt are metals.\,\cite{subramanian_oxide_1983} The 
behavior of Bi$^{3+}$ is suggested to be quite different in insulating and 
metallic pyrochlores. In insulating \bto\ and Bi$_2$Sn$_2$O$_7$ (cubic
above 920\,K) the Bi is offset by $\sim$0.4 \AA\ from the ideal site. This
Bi off-centering is aperiodic, but otherwise analogous to the \emph{correlated}
motion of lone-pair active Bi$^{3+}$ in the ionic conductor
Bi$_2$O$_3$ \cite{norberg_comparison_2010} or multiferroic BiFeO$_3$.
\cite{hill_density_2002} In metallic pyrochlores, the suggestion is that 
weaker $A$--O$^\prime$ interactions preclude any displacement at
all.\cite{walsh_electronic_2006,hinojosa_first-principles_2008,hinojosa_influence_2010}
However, Rietveld refinements for compounds where $B$ = Ru, Rh, and Ir show 
that Bi off-centering is still present in experiment.\cite{kennedy_oxygen_1996,kennedy_structural_1997,avdeev_static_2002,tachibana_electronic_2006}
The short-range correlation of Bi displacements has been probed using 
reverse Monte Carlo modeling of the diffuse streaks in electron 
diffraction patterns.\cite{goodwin_real-space_2007}

\begin{figure}
\centering\includegraphics[width=0.85\columnwidth]{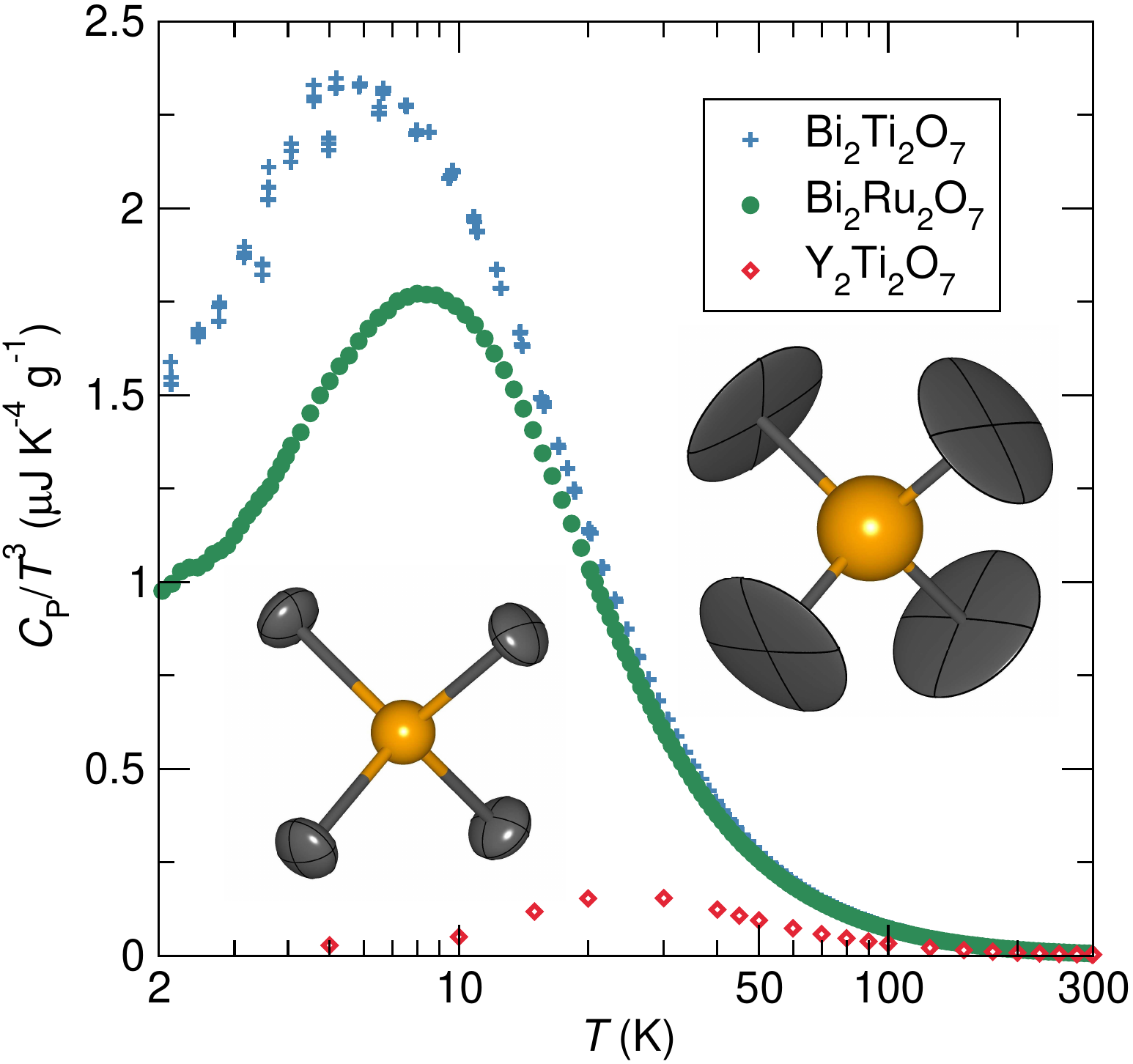} \\
\caption{(Color online) Heat capacities of three pyrochlores show a
strong dependence on the magnitude of static $A$-site disorder (large
in \bto\, present in \bro\, none in Y$_2$Ti$_2$O$_7$). Note that the
electronic contribution to the heat capacity of \bro\/ has been subtracted. 
The results of Rietveld refinement of Bragg neutron scattering displayed 
as 95\% ellipsoids on O$^\prime$ (orange) and Bi (black) on \bro\ (left) and 
\bto\ (right) point to static Bi$^{3+}$ disorder being enveloped within the 
large disks. Data adapted from 
refs.\cite{melot_large_2009,tachibana_electronic_2006,johnson_thermal_2009}.}
\label{fig:heat-capacity}
\end{figure}

In this contribution, we compare structural details in insulating \bto\ and 
metallic \bro\ using pair distribution function (PDF) analysis with 
least-squares and reverse Monte Carlo (RMC) modeling. These techniques reveal 
the precise structural tendencies of Bi$^{3+}$ off-centering, even in the case 
of incoherent ice-like disorder.\cite{shoemaker_atomic_2010} 
Specific details of bond distances, angles, and real-space shapes can be 
extracted from the RMC model because it is predicated on fits to the 
atom-atom distances in the PDF. The work should be placed in the context of 
ionic off-centering in metallic systems and the screening of ferroelectric 
dipoles, as first suggested by Anderson and Blount,\cite{anderson},  
which finds application in the context of heavily-doped perovskite 
titanates.\cite{page_prl} We find that the large displacements in 
\bto\ can be reconciled with bond valence analysis, but displacement in 
\bro\  are not driven by valence considerations alone. 

To introduce the comparison, Fig.\,\ref{fig:heat-capacity} shows that the
scaled heat capacities of \bto\/ and \bro\/ display pronounced low-temperature 
humps in plots of $C/T^3$ \textit{vs.} $T$ that are indicative of local 
Einstein modes, suggestive of glassy disorder. This large local-mode 
contribution is largely absent in Y$_2$Ti$_2$O$_7$, which has no lone pair and 
no experimentally resolvable displacive disorder.\cite{melot_large_2009}
Disorder on the O$^\prime$Bi$_4$ sub-lattice is also seen in the average 
structure Rietveld refinement of Bragg neutron scattering, masquerading as 
very large atomic displacement parameters (ADPs) on the Bi sites, displayed 
as 95\% ellipsoids in Fig.\,\ref{fig:heat-capacity}.

\section{Methods}

Preparation and average structure analysis of \bto\ powder used  in
this study has been reported by Hector and Wiggin.\cite{hector_synthesis_2004} 
\bro\ was prepared as single crystals by Tachibana \cite{tachibana_electronic_2006} 
and finely ground prior to measurement.
Time-of-flight  (TOF) neutron powder diffraction was collected at the NPDF
instrument at Los Alamos National Laboratory at 298\,K and 14\,K.
Rietveld refinement made use of the 
\textsc{GSAS} code.\cite{larson_general_2000}. Extraction of the PDF with
\textsc{PDFGetN} \cite{peterson_pdfgetn_2000} used $Q_{max}$ = 35\,\AA$^{-1}$.
Reverse Monte Carlo simulations were run using \textsc{RMCProfile} 
version 6.\cite{tucker_rmcprofile_2007} Electronic
densities of states (DOS) were calculated by the linear muffin-tin orbital
method within the atomic sphere approximation using version 47C of the
Stuttgart TB-LMTO-ASA program.\,\cite{jepsen_stuttgart_2000} Bond valence sums
(BVS) were extracted from the RMC supercell as described in previous work on
CuMn$_2$O$_4$,\cite{shoemaker_unraveling_2009} using the $R_0$ values of Brese
and O'Keeffe.\cite{brese_bond_1991}

\section{Results and Discussion}

\begin{figure} \centering\includegraphics[width=0.95\columnwidth]{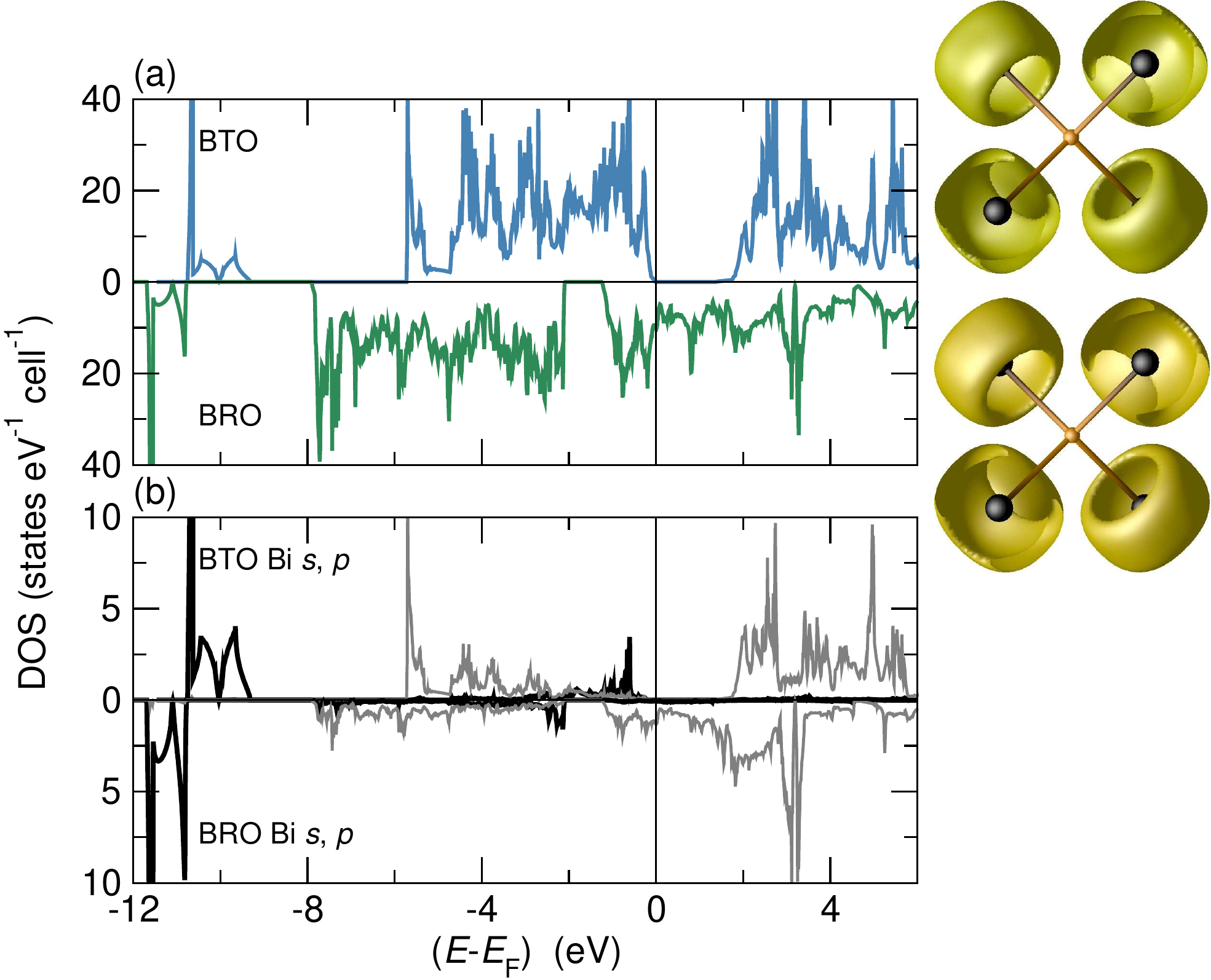} \\
\caption{(Color online) The computed total densities of states (DOS) (a,b) for
\bto\ and \bro. Below, Bi $s$ (dark) and $p$ (light) states are shown.
Electron localization functions (ELFs) for both
compounds are shown alongside as gold lobes around black Bi ions. Both
ideal structures show annuli of lone pairs around Bi ions. 
\label{fig:dos-elf}} \end{figure}

The total DOS for \bto\ and \bro\ are shown in Fig.\,\ref{fig:dos-elf}(a,b). 
The features are similar, with metallic \bro\ shifted in a nearly rigid-band
fashion by approximately 2\,eV downward, in agreement with previous 
work.\,\cite{hinojosa_first-principles_2008} Partial Bi $s$ and $p$ DOS are
shown in Fig.\,\ref{fig:dos-elf}(b,c). In both cases, some Bi $s$ states are
present at the top of the filled Bi $p$ and $B$ $d$ bands. Bi $s$ states are
plotted as electron localization function (ELF $\approx$ 0.65) isosurfaces at 
the right of Fig.\,\ref{fig:heat-capacity}(a,b). With Bi in their ideal 16$c$ 
positions, the ELFs both show essentially identical circularly
averaged lone pairs.\cite{seshadri} Assuming similar lone pair--cation 
distances, the ELFs imply that both compounds should have the same cation 
displacement in the real ground state structures.

\begin{figure} \centering\includegraphics[width=0.85\columnwidth]{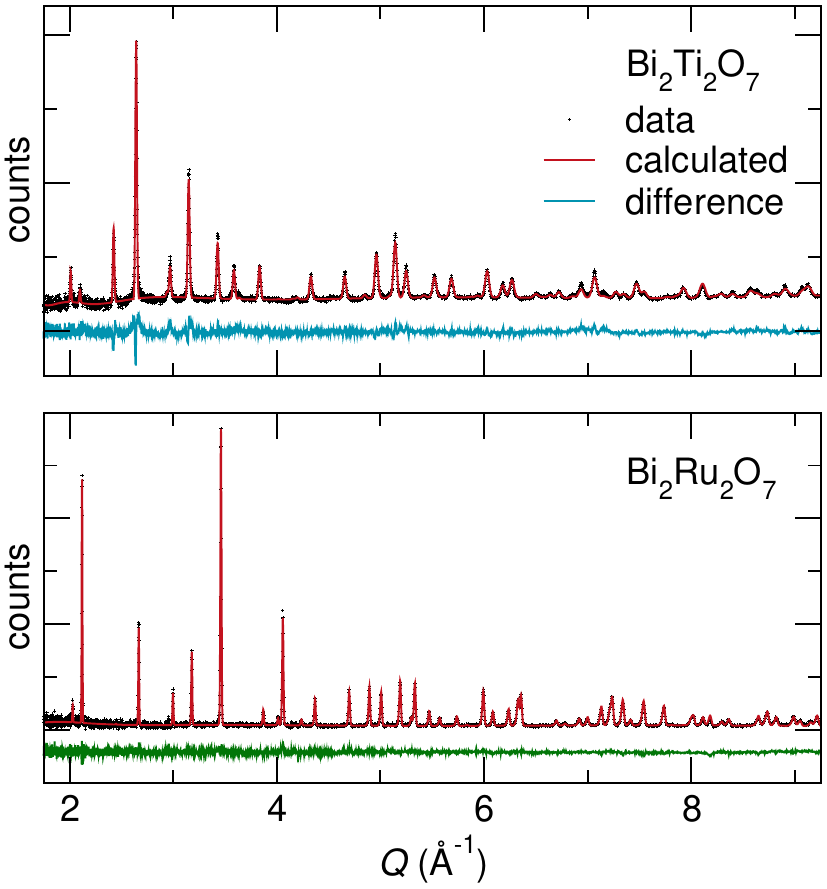} \\
\caption{(Color online) Time-of-flight neutron diffraction Rietveld refinement
of \bto\ (top) and \bro\ (bottom) at 14 K using the ideal pyrochlore structure
with anisotropic thermal parameters. The fit to \bto\ is visibly worse due to 
large amounts of diffuse shoulder intensity. This diffuse scattering is a result
of large local Bi and O$^\prime$ displacements.
\label{fig:rietveld}} \end{figure}

Time-of-flight neutron diffraction Rietveld refinements at $T$ = 14\,K and 
300\,K were performed using the ideal pyrochlore model with Bi on the 
16$c$ sites and anisotropic ADPs. Fits at 14\,K are shown in 
Fig.\,\ref{fig:rietveld}. No substantial differences were found from 
the analysis of Hector and Wiggin\cite{hector_synthesis_2004} or Tachibana 
\textit{et al.}\cite{tachibana_electronic_2006} including the occupancy;  
stoichiometric \bto\ and a Bi occupancy of 0.97 for \bro.
The ADPs from 14\,K Rietveld refinement are displayed as 95\%
ellipsoids in Fig.\,\ref{fig:heat-capacity}. The most apparent
difference is the larger, disk-shaped ellipsoid representing Bi in \bto. 
Bi is known to be off-centered from the ideal 16$c$ position in a ring normal 
to the \obo\ bond.\,\cite{radosavljevic_synthesis_1998,hector_synthesis_2004,shoemaker_atomic_2010}
The large anisotropic ADPs of Bi envelop this ring. A split Bi
position can give a better fit to the diffraction data. The previous study
found that there is a slight tendency for Bi to prefer the 96$h$
positions.\cite{shoemaker_atomic_2010} This represents a six-fold splitting of
the Bi into sites that are displaced $\sim 0.4$ \AA\ from the ideal site, and
pointing \emph{between} nearby O ions on 48$f$ sites. Bi ADPs in \bro\ also
suggest displacive disorder. They too are anisotropic and 
appear as slightly flattened ellipses in Fig.\,\ref{fig:heat-capacity}.

\begin{figure} \centering\includegraphics[width=0.95\columnwidth]{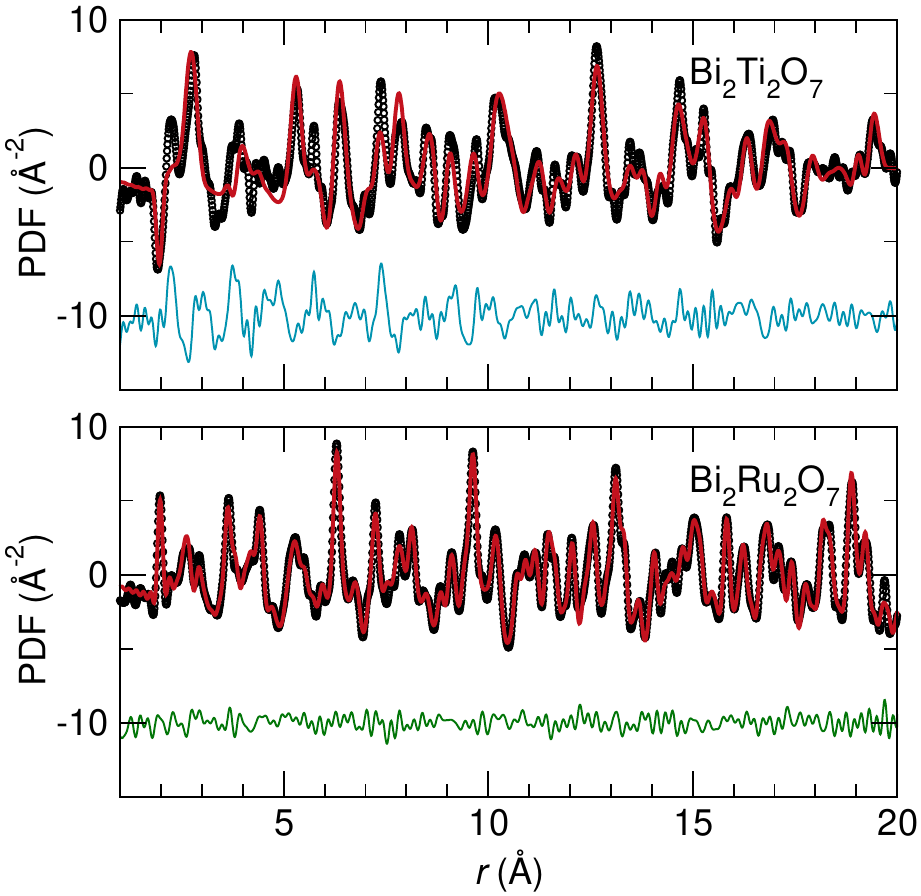} \\
\caption{(Color online) Least-squares PDF fits (using the unit cell from
Rietveld refinement) do not reproduce the low-$r$ structure of \bto\ due
to the inability of a unit-cell based description to accommodate incoherent static
displacements. The fit to \bro\ is significantly better because the
static displacements are smaller.
\label{fig:pdfgui}} \end{figure}

The most straightforward way to compare the Rietveld-refined unit cell
with the local structure is via least-squares PDF refinements,
shown in Fig.\,\ref{fig:pdfgui}. Two issues
should be considered. First, the fit for \bto\ is significantly worse overall
than \bro. This implies that the local structure of \bto\ is more poorly 
described by the $Fd\overline{3}m$ unit cell. Second, the fit for
\bto\ is poorest at low $r$, which contains details of
near-neighbor atomic distances (of particular importance are
Bi--O and Bi--O$^\prime$), and is still unsatisfactory 
at higher $r$ even though a larger number of pairs are being included.
The fit for \bro\ is equally decent at all $r$ values up to 20 \AA.

These PDF fits do not give the positions of atoms
in either compound, but 
they quickly reveal valuable information about the \emph{presence} of 
local atomic displacements (more apparent in \bto\ than \bro) and the
correlations between them (still unable to be averaged for $r < 20$ \AA).
Extracting structural tendencies of geometrically
frustrated compounds via least-squares refinement is inherently difficult
because there is no straightforward
way to model the large, complex collection of discrete displacements needed to
reproduce the disorder.
Least-squares is not a suitable
algorithm for determining so many free Bi positions, especially when their
interactions may be correlated. Instead, we remove symmetry constraints and
use RMC to investigate how
Bi are distributed within a large supercell. 

\begin{figure} \centering\includegraphics[width=0.95\columnwidth]{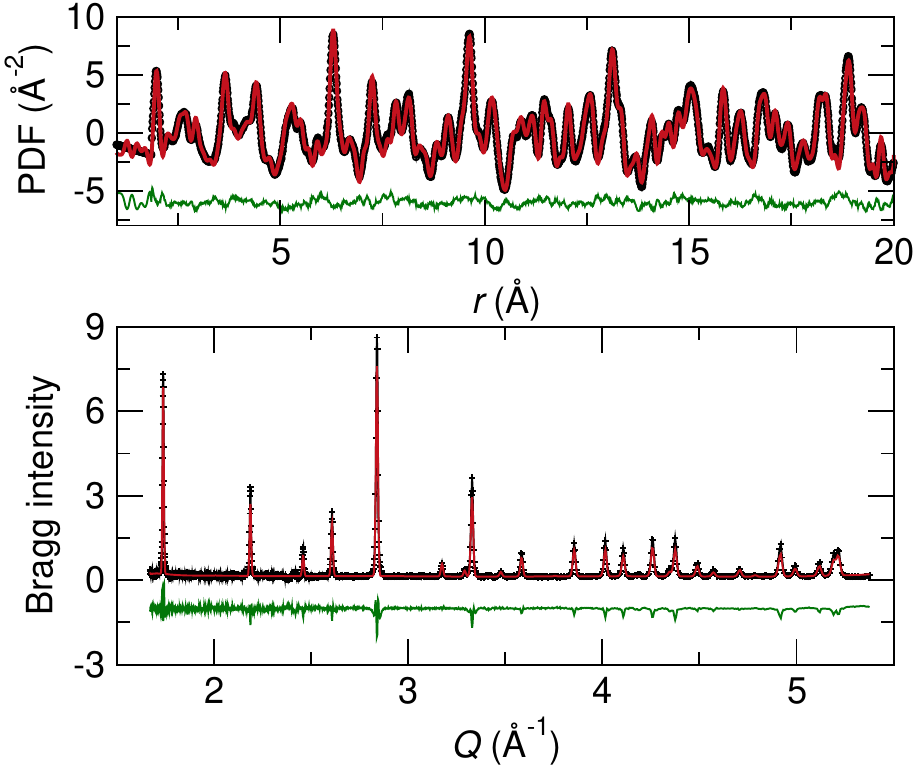} \\
\caption{(Color online) RMC fits to the PDF and Bragg profile of \bro\ at
14 K constrain the local structure (atom--atom distances) and long-range
periodicity (cell parameter, ADPs)
\label{fig:rmc-fits}} \end{figure}

\begin{figure}
\hspace{0.5cm} \includegraphics[width=0.85\columnwidth]{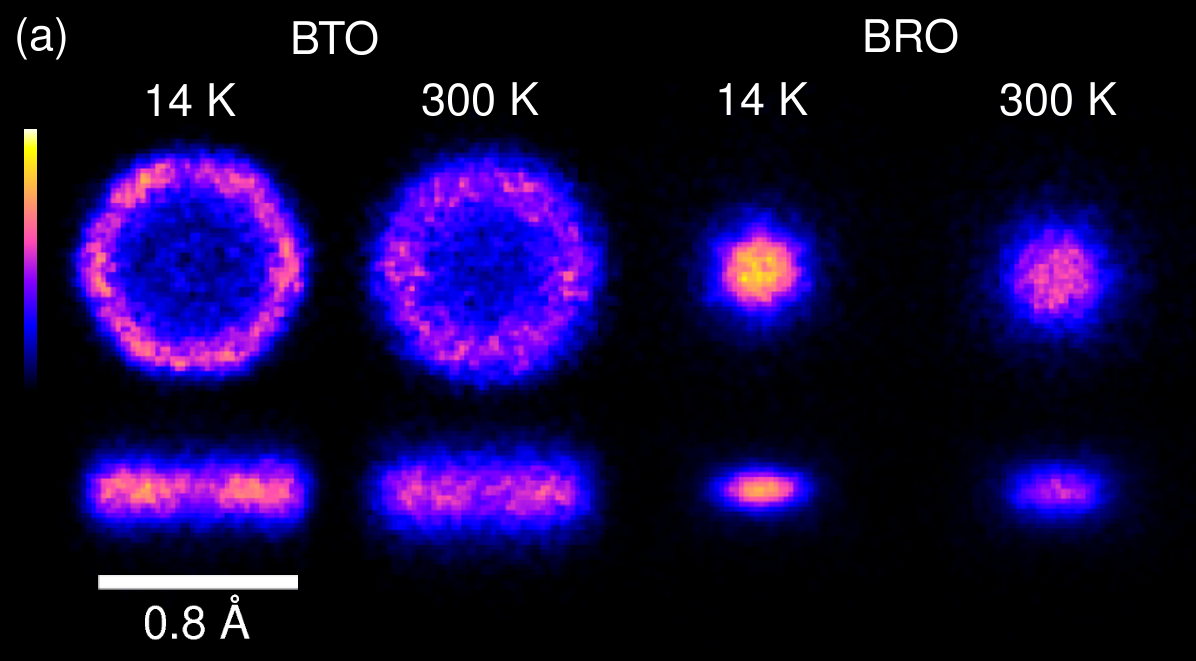} \\ \vspace{0.2cm}
\centering\includegraphics[width=0.9\columnwidth]{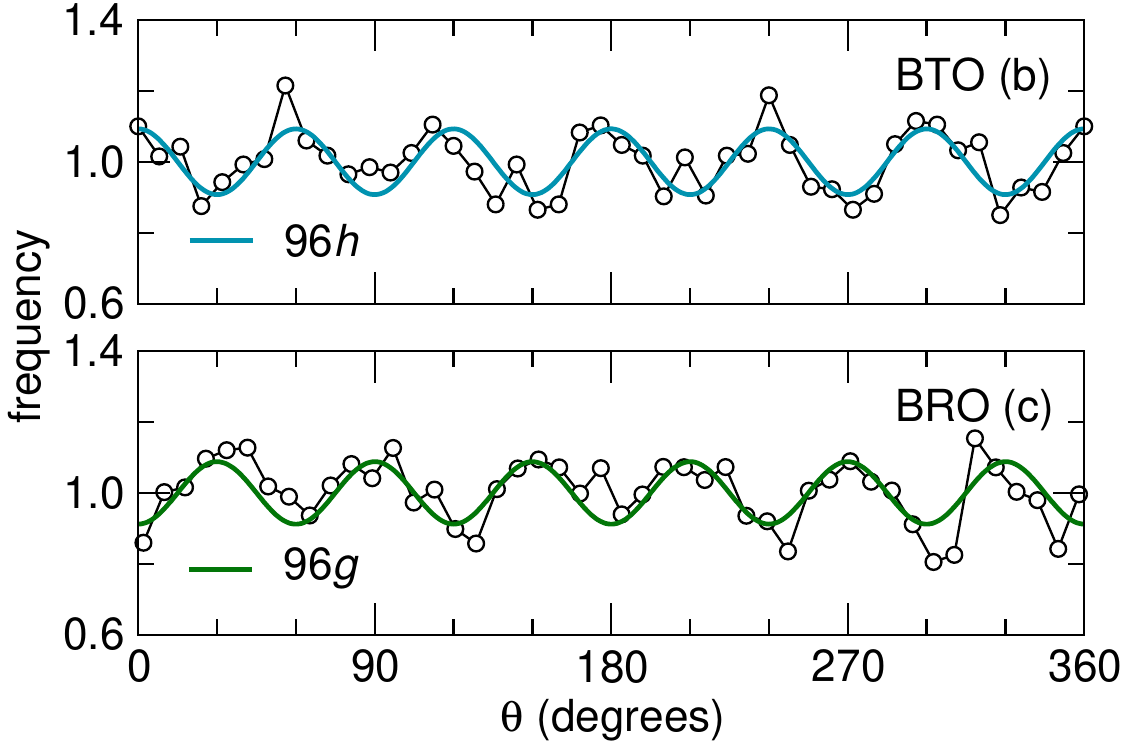} \\
\caption{ (Color online) 
Clouds of Bi nuclear intensity (a) for \bto\ and \bro\
are viewed along (top) and normal to (bottom) the \obo\ bond. Static disorder
in both produces hexagonal ring or disk shapes centered on the ideal position.
$B$-site cations Ti and Ru are (bottom) are shown for size comparison. 
Distribution of Ti and Ru as a function of angle around the ring $\theta$
is shown in (b) and (c). The sixfold modulations indicate a preference
for $96h$ and $96g$ sites. Data in (b) adapted from
ref.\,\cite{shoemaker_atomic_2010}.}
\label{fig:clouds}
\end{figure}

Simulations were carried out 
using the RMC method to investigate the precise positions of Bi. Simultaneous
fits to the PDF and Bragg profile for \bro\ are shown in Fig.\,\ref{fig:rmc-fits}.
Unit-cell based modeling (least-squares refinements, including Rietveld) usually fails
to model incoherent static displacements. In contrast,
each RMC supercell contains thousands of ions of each type. 
Folding the RMC supercell into a single unit cell produces ``point clouds'' 
of ions at each crystallographic site. These clouds display the propensity of
ions to displace from their ideal positions. The mean squared
displacement of points are in quantitative agreement with the  average ADPs
obtained from Rietveld refinement. Mapping these points as two-dimensional
histograms (Fig.\,\ref{fig:clouds}) shows the tendency of Bi nuclei to offset
in \bto\ and \bro. The Bi clouds are viewed normal to the \obo\  bond (top)
and perpendicular (bottom) for two temperatures.

The Bi ring in \bto\ is evident from Fig.\,\ref{fig:clouds} and has a
diameter of $\sim$0.8 \AA.  At 300\,K, the ring is more diffuse. We
attribute this to thermal broadening. Interestingly, the ring is not a perfect 
circle. It has a sixfold symmetry corresponding to the preference 
for Bi to occupy the 96$h$ positions (corners of the hexagon), which point 
between the six neighboring 48$f$ O ions in the 
TiO$_6$ network.\cite{shoemaker_atomic_2010}

The Bi distribution in \bro\ is distinctly different, but static
displacement is still present. The displacements are densely clustered close 
to the ideal position and there is no hollow center as in \bto. However, the 
perpendicular view reveals that the Bi distribution is still disk-shaped.
Most surprising is the symmetry of the disk. It also has a hexagonal shape 
but the hexagon is rotated 30$^\circ$ with respect to what is seen in \bto,
with flat edges on the left and right, and corners on the top and bottom. This 
implies that Bi is displacing \emph{toward} the nearby 48$f$ O onto $96g$ 
positions. The Bi offset roughly agrees with the value of 0.16\,\AA\ found 
in the split-site model of Avdeev.\,\cite{avdeev_static_2002} 

Quantitative RMC Bi nuclear density as a function of the angle
$\theta$ around a ring normal to the \obo\ bonds is shown in
Fig.\,\ref{fig:clouds}(b,c). Both compounds show sixfold modulation fit by a
cosine curve with a period of 60$^\circ$, but their oscillations are offset by
30$^\circ$.

\begin{figure}
\centering\includegraphics[width=0.85\columnwidth]{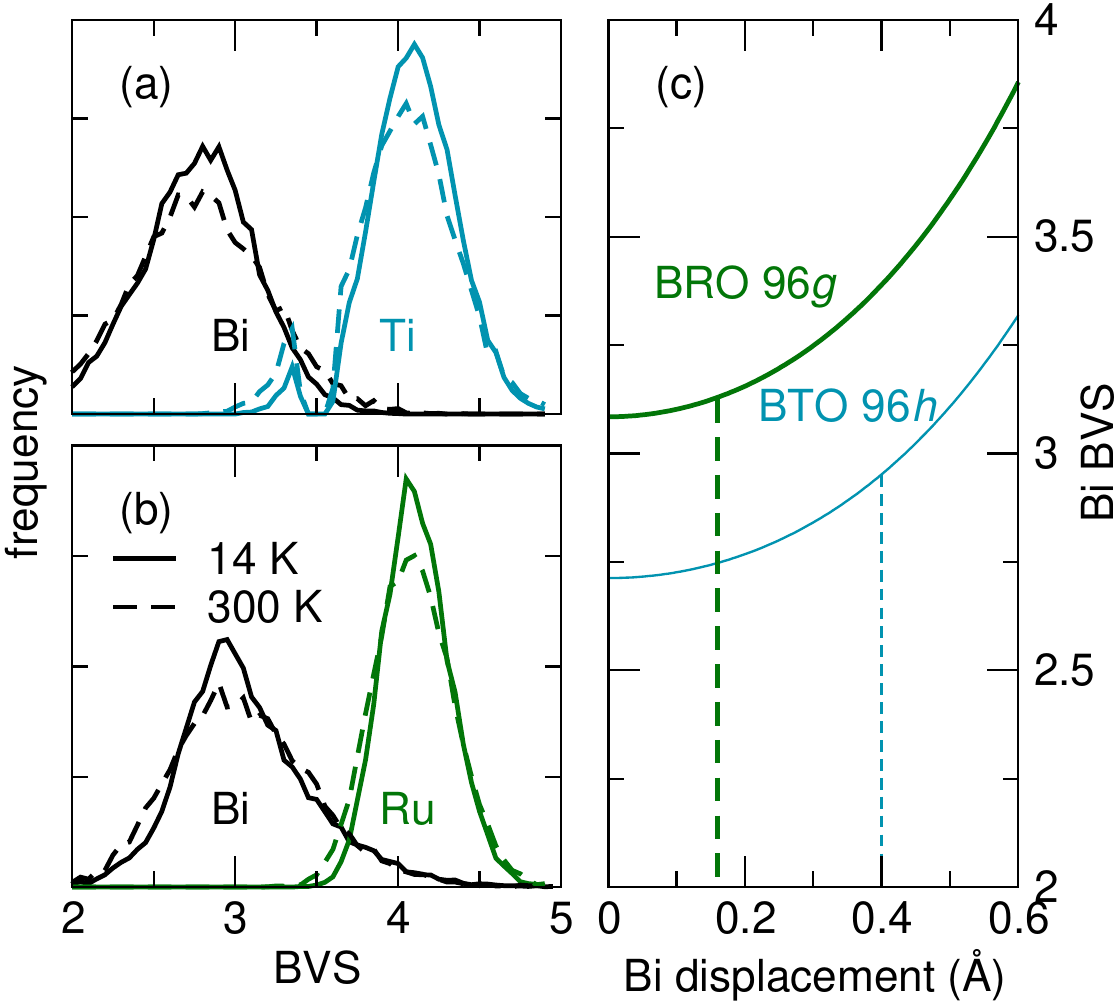}
\\ \caption{ (Color online) Bond valence sums (BVS) extracted from the RMC
super-cells show entirely Bi$^{3+}$ in
(a) \bto\ and (b) \bro. The small shoulder of Ti$^{3+}$ is a consequence of
peak broadening. Only Ti$^{4+}$ (note peak center) exists in the
sample. In (c), Bi BVS is plotted as a function of displacement into 96$g$ and
96$h$ positions. \bto\ requires significant off-centering to obtain valence
from 48$f$ O anions. Dashed lines show the actual displacement and BVS.
} \label{fig:bvs} \end{figure}

Bond valence sums are calculated for each individual cation in the supercell
and plotted as histograms in Fig.\,\ref{fig:bvs} for (a) \bto\ and (b) \bro\
at $T$ = 14 and 300 K. In all cases, BVS distributions are centered on the
expected valence: Bi$^{3+}$, Ti$^{4+}$, Ru$^{4+}$, and O$^{2-}$. A slight
sharpening is seen for the low-temperature measurement. These distributions
reveal that the RMC simulations contain chemically reasonable bond lengths 
despite the absence of such distance constraints in the simulations. They also 
reveal that there is no tendency for Bi$^{5+}$ in \bro, supporting
the conclusion from LMTO calculations that Bi $5s$ states are localized
far below the Fermi energy.

The calculated Bi BVS
versus displacement in Fig.\,\ref{fig:bvs}(c) demonstrates why displacements
are more pronounced in \bto. In
the average structures, \bto\ and \bro\ respectively obtain only 1.31+
and 1.38+ per Bi from bonds to O$^\prime$. The majority of the valence is obtained
from 48$f$ O and increases with Bi displacement, represented by the upward
curve. 
%BVS suggests no distinction in valence
%between 96$g$ or 96$h$ sites in either compound.
Bi in both compounds gain about the same valence from
48$f$ O but a large displacement is required to reach the retracted
Ti$_2$O$_6$ sublattice. The result is that each Bi in \bto\
gains an uneven amount of charge
from the six 48$f$ O, locking in dipoles.
This explains why the Bi displacements appear to be static in variable
temperature measurements.\cite{shoemaker_atomic_2010}
The rigidity of the $B_2$O$_6$ sublattice is confirmed by
small ADPs on 48$f$ O in both compounds, even in the presence of nearby Bi
offcentering. In \bro, the Ru$_2$O$_6$ network pushes
48$f$ O closer to Bi so that valence is satisfied. 

We have considered whether covalency could lead to difficulties in
using BVS to judge valence in \bro : should more covalent bonding 
(shorter $M$--O bonds) lead to Bi and Ru requiring
\textit{more} than the formal 3+ and 4+ to be satisfied? This
seems unlikely, not only because Rietveld refinement and RMC find
the desired states centered near the nominal
values. The average structure of semiconducting
BiCaRu$_2$O$_7$\,\cite{kennedy_preparation_1995}
displays much higher BVS sums (3.25+ for Bi, 4.20+ for Ru) than \bro,
but large anisotropic ADPs on the $A$ site portend static disorder nonetheless.

In \bto, off-centering helps satisfy Bi valence and the lone pair can be
accommodated in the opposite direction. In \bro, no off-centering is necessary
from valence considerations, so static disorder may be driven by lone-pair 
activity. Metallic screening in \bro\ is expected to decrease repulsions of 
the lone pair from nearby O,\cite{kennedy_structural_1997}
allowing Bi to stay closer to the ideal position than the traditional 
cation--lone pair distance would dictate. This is supported by the idea
that lone pairs often exhibit decreased
volumes.\,\cite{stoltzfus_structure_2007}

\section{Conclusions} 

In conclusion, reverse Monte Carlo stuctural analysis using total neutron 
scattering provides a detailed view
of the incoherent static displacements in \bto\ and \bro.
Real-space maps of static displacements reveal the distinct 
magnitudes and directions of Bi off-centering in \bto\ and \bro.
While static displacements in the insulator \bto\ can be understood 
on the basis of valence satisfaction alone, the cause for displacements in 
metallic \bro\ is not captured by first-principles calculations on the 
ideal compound or by the bond valence sum. An incoherent 
lone-pair driven distortion is present but is partially screened
by the conduction electrons.

\section{Acknowledgments}

We thank Anna Llobet, Thomas Proffen, Joan Siewenie, Katharine Page, and
Graham King for helpful discussions and their hospitality while DPS was
visiting the Lujan Center. This work utilized NPDF at the
Lujan Center at Los Alamos Neutron Science Center, funded by the DOE Office of
Basic Energy Sciences and operated by Los
Alamos National Security LLC under DOE Contract DE-AC52-06NA25396. 
Simulations were performed on
the Hewlett Packard QSR cluster at the California NanoSystems Institute.
The UCSB-LANL Institute for Multiscale Materials Studies, the 
National Science Foundation (DMR 0449354), and the use of 
MRL Central Facilities, supported by the MRSEC Program of the NSF 
(DMR05-20415), a member of the NSF-funded Materials Research
Facilities Network (www.mrfn.org) are gratefully acknowledged.
Work at Argonne National Laboratory is supported by UChicago Argonne,
a U.S. DOE Office of Science Laboratory, operated under Contract 
No. DE-AC02-06CH11357.

\bibliography{bto7-bro7}

\end{document}